\documentclass[11pt]{article}
\usepackage{moriond,epsfig}

\def\be{\begin{equation}}
\def\ee{\end{equation}}
\def\bea{\begin{eqnarray}}
\def\eea{\end{eqnarray}}

\begin{document}
\vspace*{4cm}
\title{HIGH-ENERGY NEUTRINO ASTRONOMY: STATUS AND PROSPECTS FOR COSMIC RAY PHYSICS}

\author{ V. VAN ELEWYCK }

\address{APC, Universit\'e Paris Diderot, CNRS/IN2P3, CEA/Irfu, Obs. de Paris, Sorbonne Paris Cit\'e, France}

\maketitle\abstracts{
Neutrino astronomy has entered an exciting time with the completion of the first km$^3$-scale neutrino telescope at the South Pole (IceCube) and the successful operation of the first undersea neutrino telescope in the Mediterranean (Antares). This new  generation of experiments is approaching the sensitivity levels required to explore at least part of the current landscape of neutrino flux predictions from astrophysical sources, bringing neutrino astronomy on the verge of its first discovery. This contribution presents the current status and latest results of the operating neutrino telescopes, with a particular emphasis on the link with the phenomenology of high-energy cosmic rays.}

\section{High-energy neutrino astronomy: motivations and instruments}

Neutrinos have long been proposed as an alternative to cosmic rays (CRs) and photons to explore the high-energy sky, as 
they can emerge from dense media and travel across cosmological distances without being deflected by magnetic fields nor absorbed by ambient matter and radiation~\cite{becker,chiarusi,baret,katz}.  Astrophysical high-energy ($\sim$TeV-PeV) neutrinos are thought to originate from  the decay of $\pi$ and $K$ produced in the interactions of accelerated protons and nuclei with matter and radiation in the vicinity of the source.  Candidate neutrino emitters therefore include most known or putative cosmic ray accelerators, ranging from galactic sources such as supernovae remnants or microquasars to the most powerful extragalactic emitters such as Active Galactic Nuclei (AGNs) and Gamma-Ray Bursts (GRBs)~\cite{becker}. Apart from unambiguously tracing the existence of hadronic processes inside known classes of astrophysical objects, the observation of cosmic neutrinos could also reveal new types of sources, yet unobserved with photons or CRs.

At even higher energies ($\sim$ EeV), one also anticipates the existence of a diffuse flux of neutrinos produced by the interaction of ultra-energetic cosmic rays (UHECRs, observed up to $E\sim 10^{20}$ eV), with the diffuse photon backgrounds filling the Universe, through the Greisen-Zatsepin-Kuzmin mechanism~\cite{GZK}.  Depending on the composition of the UHECRs, this `cosmogenic' (or GZK') neutrino flux could also come within observational reach of the instruments dedicated to neutrino astronomy, especially if combined with other detection techniques such as radio antennae (see~\cite{baret} for a short discussion about UHE neutrino detection techniques). Due to oscillations, the expected flavor ratio at Earth for both cosmic and GZK neutrinos is $\nu_e : \nu_\mu : \nu_\tau = 1 : 1 : 1$.

Neutrino astronomy is challenged both by the weakness of neutrino interactions and by the feebleness of the expected fluxes of cosmic neutrinos, which complicates their identification among the abundant background of terrestrial neutrinos produced in CR-induced atmospheric showers.   High-energy neutrino telescopes, whose detection principle was first suggested by Markov in the 60's~\cite{markov}, are designed to detect the charged leptons produced when a neutrino interacts with material around the detector. The Cherenkov light emitted by these leptons when they travel in a transparent medium such as water or ice can be detected by a 3-dimensional  array of photosensors. Muons are the preferred detection channel, but showers induced by electron- and tau-neutrino can also be detected. The timing, position and amplitude of the hits recorded by the photosensors allow the  reconstruction of the muon trajectory, providing the arrival direction of the parent neutrino and an estimation of its energy. Such detectors are installed at great depths and optimized to detect up-going muons produced by neutrinos which have traversed the Earth, in order to limit the background from down-going atmospheric muons. 

Three neutrino telescopes are currently operating worldwide; the most advanced one is IceCube~\cite{icecube}, which has recently achieved its final configuration with 86 strings, instrumenting one km$^3$ of South Pole ice. Each string supports 60 optical modules (OMs) enclosing one photomultiplier each, with its electronics board. IceCube possesses a denser infill of 6 additional strings dubbed as DeepCore~\cite{deepcore}, which extends its detection capabilities at lower energies (down to $\sim$ 10 GeV). Each of the IceCube strings is also complemented on the surface by two Cherenkov tanks filled with frozen water and overlooked by two OMs, forming a 1 km$^2$ air shower array named IceTop~\cite{icetop}. It is used for CR composition studies, and in coincidence with the in-ice detector.  Another neutrino telescope has been operating for some years in Lake Baikal~\cite{baikal} in a much smaller configuration; it has deployed since 2008 2 prototype strings for a km$^3$-scale detector. Finally, ANTARES~\cite{antares} is a neutrino telescope deployed in the Mediterranean Sea near Toulon (France), at a depth of about 2,5 km.  It is operating in its complete configuration since 2008, with 885 photomultipliers enclosed in OMs  and distributed in triplets on 12 detection lines. It is currently the largest neutrino telescope in the Northern Hemisphere; its location allows for surveying a large part of the Galactic Plane, including the Galactic Centre, thus complementing
the sky coverage of IceCube. 


\section{Searches for cosmic neutrinos}

- {\bf Diffuse flux searches} 

\vspace*{0,2cm}

\begin{figure}[t]
\hspace*{-0.5cm}
\includegraphics[width=8cm,height=6cm]{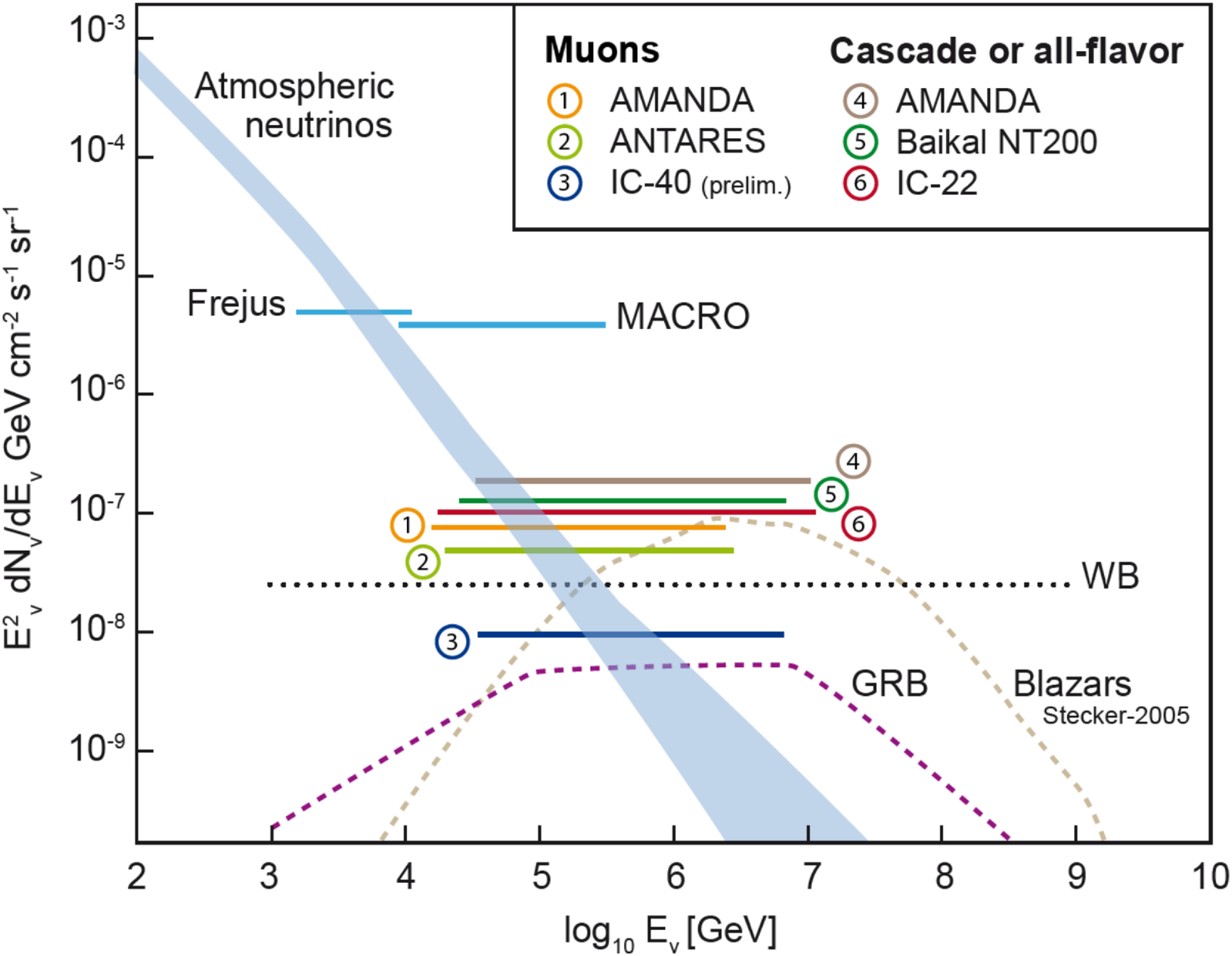}
\hfill
\includegraphics[width=8cm,height=6cm]{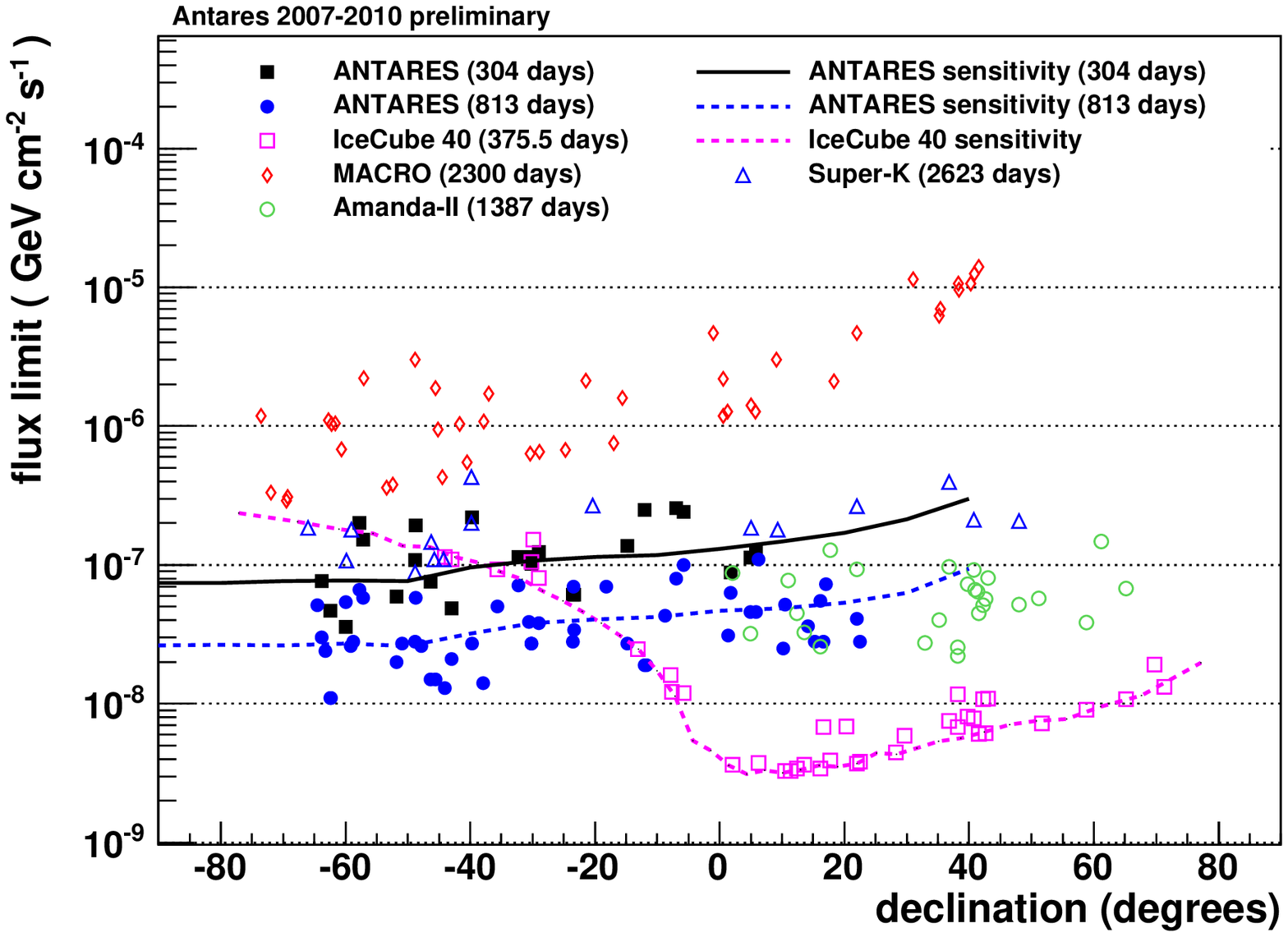}
\caption{\footnotesize {\bf Left:} 90\% C.L. integral upper limits on the diffuse flux of extraterrestrial neutrinos (normalized to one flavour), as taken from~\protect\cite{katz}, where all references can be found. The coloured
band indicates the measured flux of atmospheric neutrinos.  {\bf Right:} 90\% C.L.  upper limits for a neutrino flux with an $E^{-2}$ spectrum for candidates
sources (points) and the corresponding sensitivities (lines) in function of declination as obtained with ANTARES and IceCube~\protect\cite{antaresICRC}. The figure also shows the results of the underground MACRO and Super-Kamiokande experiments (which were not principally devoted to neutrino astronomy).  }
\label{astro}
\end{figure}

These searches look for an all-sky excess of high-energy events above the irreducible background of atmospheric neutrinos.  Such methods are especially suited to reveal the bulk of extragalactic sources (such as active galactic nuclei or gamma-ray bursts) which are too faint to be detected individually. They rely on energy estimators which are typically related to the number and multiplicity of photon hits in the PMTs.  No significant excess over atmospheric neutrinos has been
observed so far. An overview of the limits set in the TeV-PeV range by ANTARES, IceCube and their predecessors, using either muon tracks (induced by $\nu_\mu$ only) or contained showers (induced by all flavors of neutrinos), are presented in Fig.~\ref{astro} (left). The most stringent one~\cite{ice3diff} (from IceCube in its 40-string configuration) provides an upper limit on the normalization of an $E^{-2}$ astrophysical $\nu_\mu$ flux of  $8,9\ 10^{-9}$ GeV cm$^{-2}$ s$^{-1}$ sr$^{-1}$, which already excludes the most optimistic models of neutrino production in astrophysical sources. This sensitivity level also exceeds 
the Waxman-Bahcall bound~\cite{wb} which has long served as a benchmark flux for neutrino telescopes (and would provide $\sim $100 events per year in a km$^{3}$ detector).

Above the PeV, the sky coverage of neutrino telescopes is reduced because of neutrino absorption in the Earth. It can be partially recovered by looking for horizontal and down-going neutrinos, at the price of significantly increasing the energy threshold of the detector. A limit on the neutrino flux in the PeV-EeV range  has been recently set by IceCube~\cite{ice3gzk}, which is competitive with the ones set by radio and air shower experiments. 

\newpage

\noindent
- {\bf Point source searches} 

\vspace*{0,2cm}

These searches aim at detecting significant excesses of events from particular spots (or regions) of the sky. They can be performed either blindly over the full sky, or in the direction of a-priori selected candidate source locations that correspond e.g. to known gamma-ray emitters. Such searches are well-suited to look for steady, point-like sources, in particular in the Galaxy. They rely on a good angular resolution, which above 10 TeV is essentially determined by the scattering length of light in the medium, yielding a median error angle on the neutrino direction of about $0.5^{\circ}$ for IceCube and $0.4^{\circ}$ for ANTARES (note that one can expect  $0.1^\circ$ resolution for a km$^3$ deep-sea detector). The latest results of point sources searches by ANTARES and IceCube (in its 40-string configuration) are presented in Fig.~\ref{astro} (right), showing the nice complementarity in their field of view and the large improvement in sensitivity gained over the past 15 years by the development of dedicated instruments for neutrino astronomy.

\vspace*{0,2cm}

\noindent
{\bf - Multimessenger programs} 

\vspace*{0,2cm}

Neutrino telescopes also develop specific strategies to look for neutrinos with timing and/or directional correlations with potential sources of gamma-rays (in connection with the GCN alerts), UHECRs (such as CenA) and gravitational waves. The limited space-time search window allows a drastic reduction of the atmospheric background, therefore enhancing the sensitivity to faint signals that would have remained undetected otherwise. This method has been used e.g. to put upper limits on the total (stacked) flux from samples of GRBs that occurred while IceCube was taking data. The most stringent of these limits, obtained with IceCube 40 and 59 data, now questions the fireball paradigm of UHE proton acceleration in GRB, and hence the very fact that these sources be an important contributor to the UHECR flux~\cite{ice3grb}. 

Alternatively, the occurrence of a special event in a neutrino telescope (such as the near-simultaneous arrival of two or more neutrinos from the same direction) could indicate that a highly energetic burst has occurred and may be used as a trigger for optical, x-ray, and gamma-ray follow-ups.  IceCube  and ANTARES  currently have alert programs established or in development with e.g. fast optical telescope networks like ROTSE and TAROT, and gamma-ray telescopes such as Swift, Fermi, MAGIC, and VERITAS.

\section{Atmospheric neutrinos}

A crucial ingredient for performing efficient neutrino astronomy is the precise knowledge of the atmospheric neutrino and muon fluxes, which constitute the bulk of events detected in neutrino telescopes. $\nu_e$'s, $\nu_\mu$'s and muons are produced in the air showers induced by the interaction of CRs in the atmosphere. Up to about 100 TeV, the dominant production channels are $\pi^\pm$ and $K^\pm$ decays, yielding an energy spectrum $\sim E^{-3,7}$. At higher energies, an additional component originating from the prompt decay of charmed and bottomed particles takes over, with a spectrum which is expected to follow more closely that of the primary CRs ($\sim E^{-2,7}$). Its precise features are however sensitive to the hadronic interaction models, which are still poorly constrained at those high energies.

\begin{figure}[t]
\label{atmo}
\hspace*{-0.5cm}
\includegraphics[width=8cm,height=6cm]{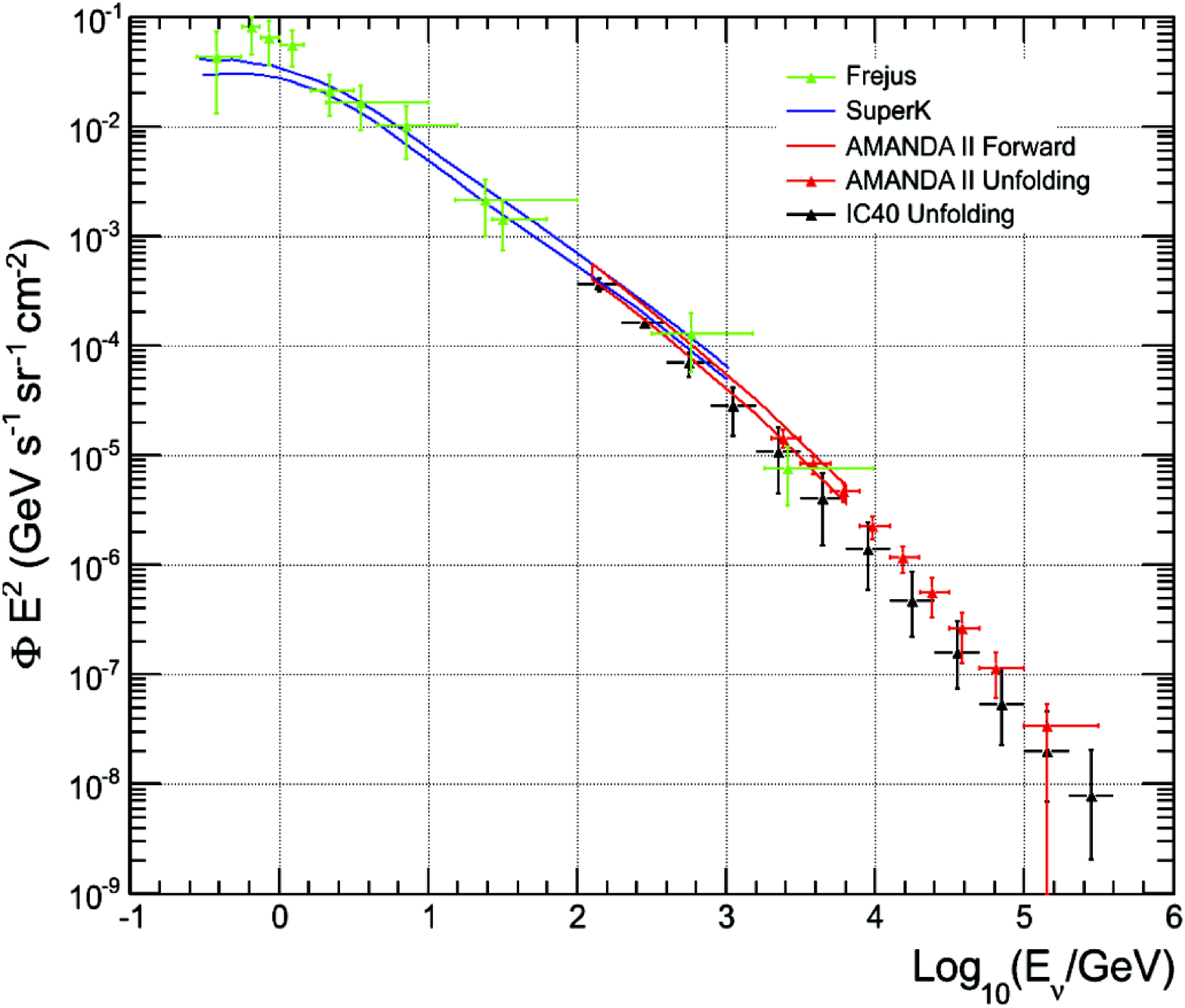}
\hfill
\includegraphics[width=8cm,height=6cm]{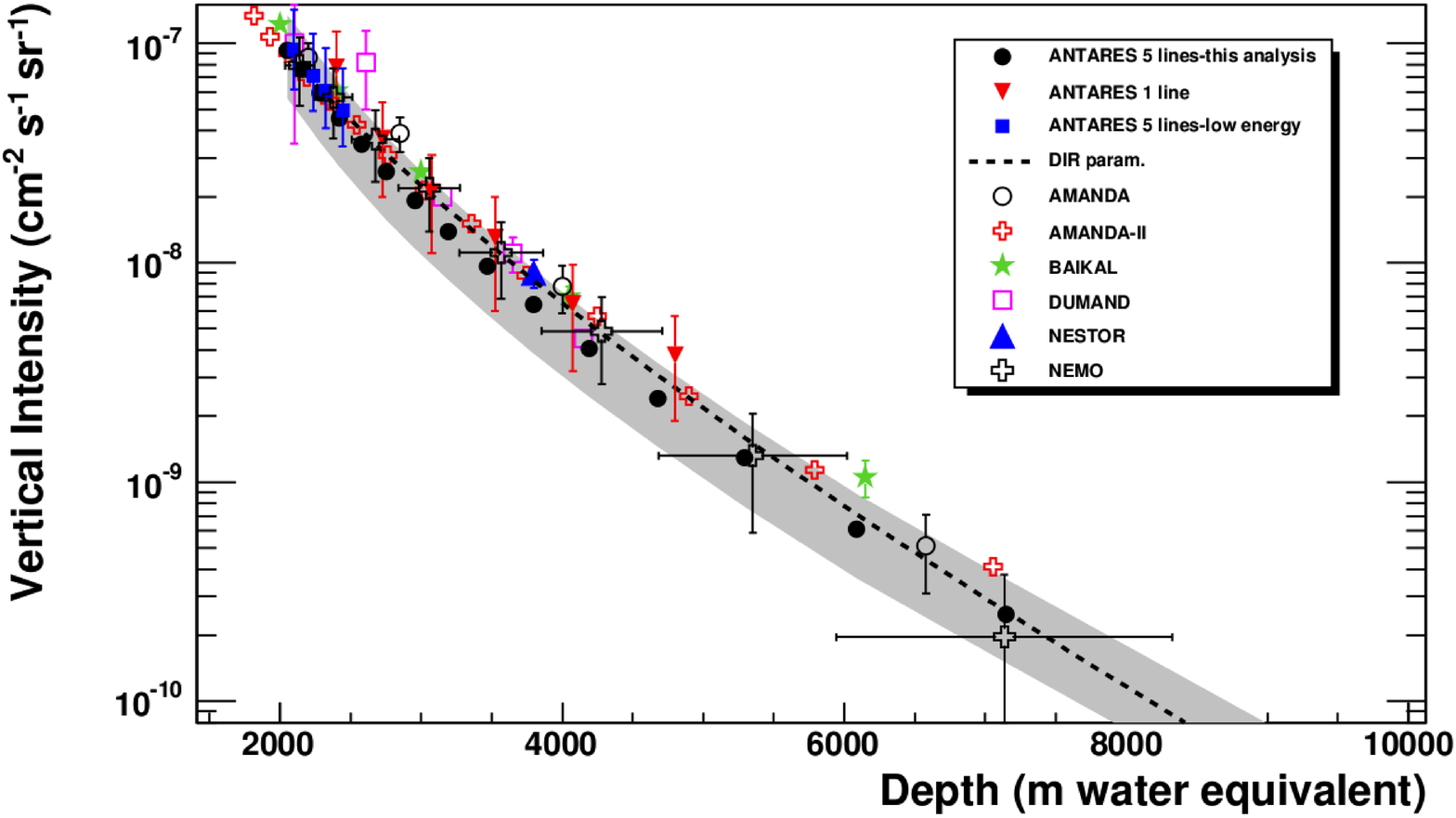}
\caption{\footnotesize{{\bf Left:} Energy spectrum of atmospheric neutrinos as observed by IceCube in its 40-string configuration, its prototype AMANDA, and previous underground experiments (SuperKamiokande, Fr\'ejus)~\protect\cite{ice3atmo}.  {\bf Right:} Depth-intensity relation for atmospheric muons as measured by various  neutrino telescopes and prototypes, the most recent one being that of ANTARES (in its 5-line configuration)~\protect\cite{antmuons} whose systematic uncertainties are represented by the shaded band. The dashed curve is the expectation derived from a popular model~\protect\cite{bugaev}. }}
\end{figure}

Neutrino telescopes have developed efficient reconstruction strategies allowing a discrimination of atmospheric $\nu_\mu$'s vs $\mu's$ on basis of the zenith angle of the track, as up-going muons can only be produced by atmospheric neutrinos that have traversed the Earth. Sophisticated unfolding algorithms are then applied to extract the energy spectrum of the parent $\nu_\mu$'s on basis of the energy deposited by the muons in the detector. The resulting spectrum of atmospheric neutrinos is presented in Fig.~\ref{atmo}, which illustrates how IceCube (and its prototype AMANDA) have significantly extended the measurements at high energies (up to $\sim$ 400 TeV) with respect to the previous Fr\'ejus and SuperKamiokande experiments~\cite{ice3atmo}. Statistics at high energies are however still too limited to allow disentangling the contribution of the prompt component.  

\section{Atmospheric muons and other cosmic ray-related studies}

Down-going, CR-induced muons provide a high-statistics sample of events which is primarily used for detector calibration studies. IceCube, for example, has been able to determine its angular resolution based on the high-significance observation of the  Moon shadow in the skymap of down-going muons~\cite{ice3moon}. The study of the atmospheric muon spectrum also provides per se interesting information on CRs physics. It is usually presented under the form of a depth-intensity relation, as illustrated in Fig.~\ref{atmo} (right), where the zenith-dependant muon flux at the detector is translated into an equivalent vertical intensity at a depth corresponding to the length of the muon path in matter. The wide aperture of neutrino telescopes allows them to scan a large range of depths; but their measurement is complicated by the difficulty of resolving the multiplicity of muon bundles originating from a single air shower, which constitutes a major source of systematic uncertainties. Novel methods for muon separation are being studied, {\it e.g.} based on the identification of TeV-scale catastrophic energy losses along the muon tracks~\cite{ice3cr}. A precise measurement of the single-muon spectrum up to several 100 TeVs would provide valuable information on the CR composition around the knee, and potential access to the prompt contribution.  With the large statistics accumulated in the last 4 years, IceCube has also been able to observe the seasonal variation of the muon rate, under the form of a $\pm 8 \%$ annual modulation highly correlated with the stratospheric temperature~\cite{ice3cr}. This measurement directly relates to the relative contribution of $\pi$'s and $K$'s to the muon production,  providing an estimate of the $K/\pi$ ratio of $0.09 \pm 0.04$ at CR median energy of about 20 TeV.
\vspace*{0,2cm}

\begin{figure}[t]
\begin{minipage}{0.58\textwidth}
\mbox{}\\
\hfill
\centerline{\includegraphics[width=\textwidth]{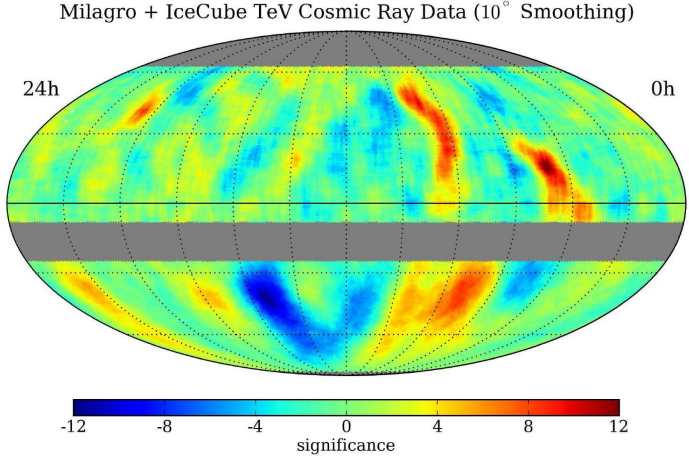}}
\end{minipage}
\hfill
\begin{minipage}{0.38\textwidth}
\mbox{}\\
\caption[spectrum]{
\footnotesize {Combined significance skymap of the distribution of cosmic ray arrival directions as observed in the Northern hemisphere by Milagro~\cite{anis} and in the Southern hemisphere by IceCube (after filtering out the dipole and quadrupole contributions)~\cite{ice3anis}. Both maps have been smoothed with a 10$^\circ$ radius.}}
\end{minipage}
\label{anis}
\end{figure}

Studies of the angular power spectrum of the atmospheric muon events recorded by IceCube have also revealed the existence of anisotropies in the arrival direction of the parent CRs above the TeV in the Southern sky, at the level of a few per mille and on several angular scales~\cite{ice3anis}. In addition to the strongest dipole and quadrupole components, the data feature several weaker regions of excess and deficit on scales between 10$^\circ$ and 30$^\circ$. Such anisotropies are of comparable (or stronger) significance as those reported so far in the Northern Hemisphere by several cosmic ray detectors, and in particular by the MILAGRO observatory~\cite{anis}, as can be seen from Fig.~\ref{anis}  (see {\it e.g.}~\cite{katz,ice3cr} for complete references). These observations challenge the common belief that the arrival directions of charged CRs at $\sim$ TeV energies should be randomized by galactic magnetic fields, and could have strong implications on our understanding of galactic CR dynamics and sources. The rapidly increasing statistics of IceCube will hopefully provide important clues to address this puzzle in the near future. 
\vspace*{0,2cm}

Finally, the combination of data from (in-ice) IceCube and IceTop provides new  opportunities for the study of primary CR mass composition~\cite{ice3cr}. The IceTop surface array is designed for the detection of air showers induced by primary cosmic rays with energies ranging between 100 TeV and 10$^{18}$ eV approximately~\cite{icetop,ice3cr}. The shower size parameter $S_{125}$ (giving the signal at a reference radius $R =125$ m from the shower axis) essentially measures the electromagnetic component of the shower, hence providing an energy proxy for the primary CR. In about 30\% of the cases, the muon bundle associated to an air shower detected by IceTop will also trigger the in-ice detector, allowing an estimation of the energy of the muon component of the shower, which is highly sensitive to the primary mass and energy.

\newpage

\section{Conclusions and perspectives}

Despite the technical difficulties associated to the operation of large-scale detectors in natural media, neutrino astronomy has witnessed tremendous instrumental progress in both hemispheres in the recent years. IceCube has now reached the km$^3$ benchmark and will hopefully be supplemented within a few years by a Mediterranean kilometric detector~\cite{km3net}, following the path opened by pioneering projects like ANTARES. Although the first evidence for a cosmic neutrino is still to come, current results from operating instruments have already started constraining our understanding of the most violent phenomena in the Universe, in particular by questioning GRBs as the dominant source of ultra-high energy cosmic rays. 

Atmospheric neutrinos and muons constitute the over-abundant physical background for cosmic neutrino searches in neutrino telescopes. They have therefore rapidly turned into a signal themselves, allowing the development of  an extended program of cosmic ray studies in the TeV to PeV range, whose first  intriguing outcome has been the observation by IceCube of large- and medium-scale anisotropies in the cosmic ray arrival directions in the Southern hemisphere. Neutrino telescopes will undoubtedly play an important role in characterizing the spectrum and composition of cosmic rays, not only up to the knee, but possibly up to $10^{18}$ eV; in this respect, the integration of IceTop and IceCube into a 3-dimensional air shower array which can be used to probe both the electromagnetic and muon components of air showers is particularly promising.

\footnotesize{
\section*{Acknowledgments}
The author would like to thank the organizers of the EDS Meeting and Rencontres du Vietnam, in particular Tanguy Pierog for inviting her to participate in this lively workshop, and Jean Tran Thanh Van for his dedication to turning Quy Nhon into such a welcoming venue for scientific events.}


\section*{References}
\begin{thebibliography}{99}
\bibitem{becker} J.~K.~Becker,
  Phys.\ Rept.\  {\bf 458} (2008) 173.
 \bibitem{chiarusi}  T.~Chiarusi and M.~Spurio,
  Eur.\ Phys.\ J.\ C {\bf 65} (2010) 649.
   \bibitem{baret} B. Baret and V. Van Elewyck, Rept. Prog. Phys. {\bf 74} (2011) 046902. 
  \bibitem{katz} U. F. Katz and C. Spiering, Prog.Part.Nucl.Phys. 67 (2012) 651.
  \bibitem{GZK} K.~Greisen,
  Phys.\ Rev.\ Lett.\  {\bf 16} (1966) 748;  G.~T.~Zatsepin and V.~A.~Kuzmin,
  JETP Lett.\  {\bf 4} (1966) 78
  [Pisma Zh.\ Eksp.\ Teor.\ Fiz.\  {\bf 4} (1966) 114].
  \bibitem{markov} M.A. Markov, Procs. Int. Conf. on High Energy Phys., p. 183, Univ. of Rochester (1960).
\bibitem{icecube} F.~Halzen and S.~R.~Klein,
  Rev.\ Sci.\ Instrum.\  {\bf 81} (2010) 081101.
  \bibitem{deepcore}  A.~Abbasi, S.~Sarkar {\it et al.} [IceCube Coll.],
  arXiv:1109.6096 [astro-ph.IM].
  \bibitem{icetop} T.~Stanev for the IceCube Collaboration,
  Nucl.\ Phys.\ Proc.\ Suppl.\  {\bf 196} (2009) 159.
\bibitem{baikal} A.~Avrorin {\it et al.} [Baikal Coll.],
  NIM\  A {\bf 626-627} (2011) S13.
  \bibitem{antares} M. Ageron {\it et al.} [ANTARES Coll.], NIM {\bf A 656} (2011) 11-38.
  \bibitem{ice3diff} R.~Abbasi {\it et al.}  [IceCube Coll.],
  Phys.\ Rev.\ D {\bf 84} (2011) 082001.
  \bibitem{wb} E. Waxman and J. Bahcall, Phys. Rev. D {\bf 59} (1999) 023002.
  \bibitem{ice3gzk}  R.~Abbasi {\it et al.}  [IceCube Coll.],
  Phys.\ Rev.\ D {\bf 83} (2011) 092003
   [Erratum-ibid.\ D {\bf 84} (2011) 079902].
   \bibitem{ice3grb}  R.~Abbasi {\it et al.} [IceCube Coll.],
 Nature {\bf 484} (2012) 351.
  \bibitem{antaresICRC} C. Bogazzi (for the ANTARES Coll.), ``Searching for Point Sources of HE Cosmic $\nu$ with the ANTARES telescope'', in Procs. 32d ICRC (2011), Beijing (China) arXiv:1112.0478.
  \bibitem{ice3atmo} R.~Abbasi {\it et al.}  [IceCube Coll.],
  Phys.\ Rev.\ D {\bf 83} (2011) 012001.
\bibitem{ice3cr} R. Abbasi {\it et al.} [IceCube Coll.], ``The IceCube Neutrino Observatory III: Cosmic Rays", in Procs. 32d ICRC (2011), Beijing (China), arXiv:1111.2735.
\bibitem{ice3moon} L.~Gladstone [IceCube Coll.],
  arXiv:1111.2969 [astro-ph.HE].
  \bibitem{antmuons} J.~A.~Aguilar {\it et al.} [ANTARES Coll.],
  Astropart.\ Phys.\  {\bf 34} (2010) 179.
  \bibitem{bugaev}  E.~V.~Bugaev {\it et al.},
  Phys.\ Rev.\ D {\bf 58} (1998) 054001.
  \bibitem{ice3anis} R.~Abbasi {\it et al.}  [IceCube Coll.],
  Astrophys.\ J.\  {\bf 740} (2011) 16.
  \bibitem{anis} A.~A.~Abdo, B.~T.~Allen {\it et al.},
  Astrophys.\ J.\  {\bf 698} (2009) 2121.
  \bibitem{km3net} P. Bagley {\it et al.} [KM3NeT Coll.], Technical Design Report (2010), available from: www.km3net.org ; V.~Van Elewyck [KM3NeT Coll.],
  PoS TEXAS {\bf 2010} (2010) 235.
\end{thebibliography}
\end{document}